\newcommand{\ket}[1]{\left|#1\right\rangle}

\documentclass[%
 reprint,
 amsmath,amssymb,
 aps,
]
{revtex4-1}

\usepackage{graphicx}% Include figure files
\usepackage{dcolumn}% Align table columns on decimal point
\usepackage{bm}% bold math

\begin{document}

%%\preprint{APS/123-QED}

\title{Measuring maximal eigenvalue distribution of Wishart random matrices with coupled lasers}

\author{Moti Fridman}
\author{Rami Pugatch}
\author{Micha Nixon}
\author{Asher A. Friesem}
\author{Nir Davidson}
\email{nir.davidson@weizmann.ac.il}
\affiliation{ Weizmann Institute of Science, Dept. of Physics of Complex Systems, Rehovot 76100, Israel}

%\author{Moti Fridman, Rami Pugatch, Micha Nixon, Asher A. Friesem, and Nir Davidson}
%\email{nir.davidson@weizmann.ac.il}
%\affiliation{ Weizmann Institute of Science, Dept. of Physics of Complex Systems, Rehovot 76100, Israel}

\date{\today}% It is always \today, today,

\begin{abstract}
We determined the probability distribution of the combined output power from twenty five coupled fiber lasers and show that it agrees well with the Tracy-Widom, Majumdar-Vergassola and Vivo-Majumdar-Bohigas distributions of the largest eigenvalue of Wishart random matrices with no fitting parameters. This was achieved with $500,000$ measurements of the combined output power from the fiber lasers, that continuously changes with variations of the fiber lasers lengths. We show experimentally that for small deviations of the combined output power over its mean value the Tracy-Widom distribution is correct, while for large deviations the Majumdar-Vergassola and Vivo-Majumdar-Bohigas distributions are correct.
\end{abstract}

\maketitle

%\tableofcontents

Random matrix theory has been exploited in numerous research fields ranging from nuclear spectra to quantum transport, models of quantum gravity in two dimensions, mesoscopic non-linear dynamics, atomic physics, wireless communications and multi-dimensional data analysis~\cite{Wigner, RevModPhys1, horizon, Mehta, websearchengines}. Of special interest are the minimal and maximal eigenvalues of random matrices, that determines for example, the conductance fluctuations in two- and three- dimensional Anderson insulators~\cite{Anderson2, Anderson3}. An analytical expression describing typical deviations of the maximal eigenvalue was presented in the 90's by Tracy and Widom (TW)~\cite{TW, TW2} initiating many further theoretical developments in random matrix theory\cite{TWpost1, TWpost2}. Recently, Majumdar and Vergassola (MV) calculated the probability of large deviations of the maximal eigenvalue~\cite{Majumdar1, Majumdar2, Majumdar3} above the mean and Vivo, Majumdar and Bohigas (VMB) calculated the probability below the mean. The MV and the VMB distributions were numerically confirmed, but so far eluded experimental demonstration.

In this letter, we provide the first experimental observation of the MV and VMB distributions in a physical system and connect the field of coupled random lasers to random matrix theory. We report our measured distribution of the combined output power from an array of 25 coupled fiber lasers whose cavity lengths randomly fluctuate in time. We found that the measured distribution of the combined output power agrees well with the distribution of maximal eigenvalue of Wishart random matrices as predicted by TW and MV. For deviations close to the mean value, the measured distribution is shown to have a universal shape that agrees with the TW distribution. For large deviations from the mean value the measured distribution deviates from the TW distribution, but agrees well with the MV and VMB distributions over more than five decades with no fitting parameters. To account for this agreement, we present a heuristic model that illustrates the relation between the output power distribution from our array of coupled lasers to the maximal eigenvalue of Wishart random matrices.

\begin{figure}[h]
\centerline{\includegraphics[width=8.3cm]{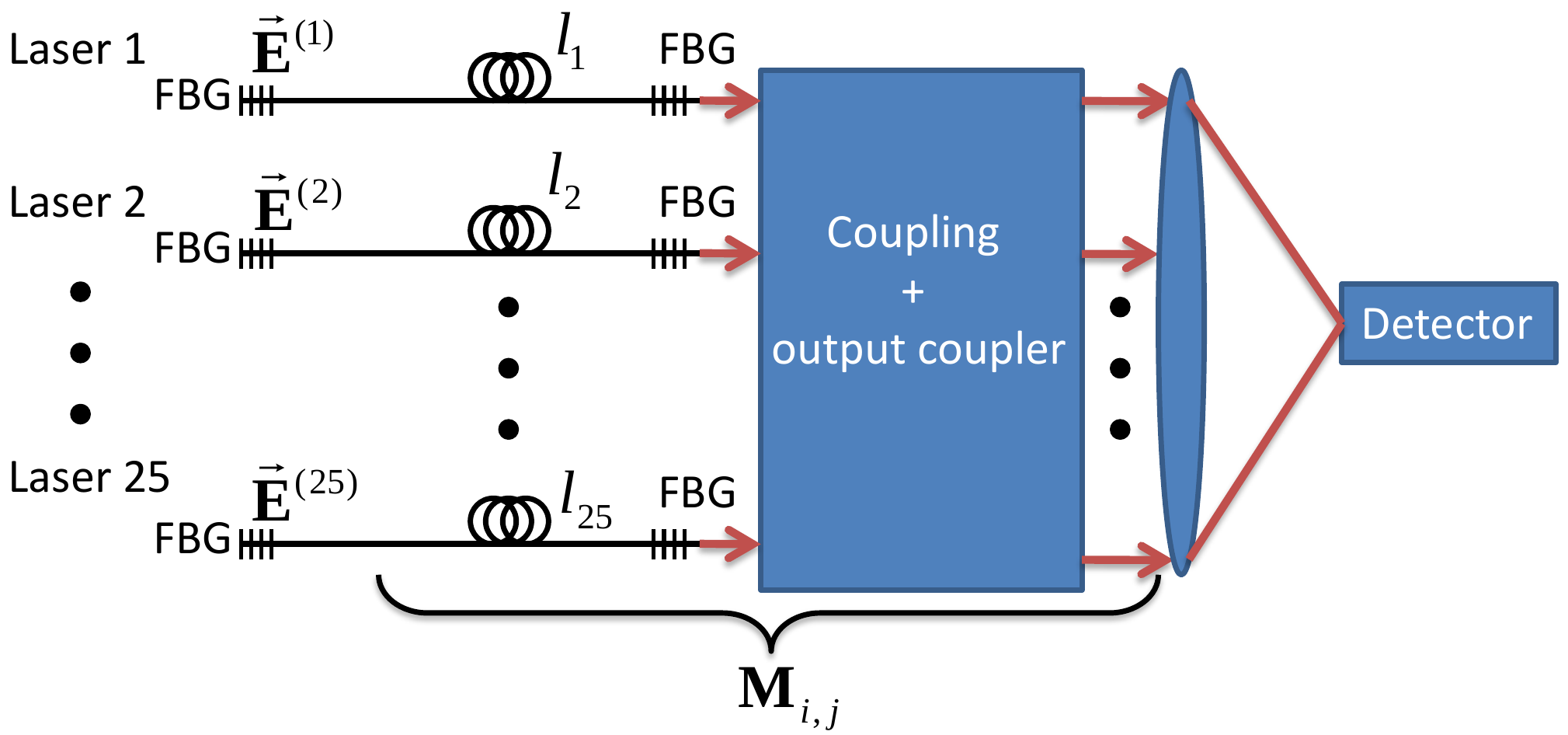}} \caption{\label{system} Experimental arrangement for measuring the combined output power distribution of 25 coupled fiber lasers. FBG - fiber Bragg gratings that serve as rear mirror ($> 99\%$ reflectivity) and front mirrors ($\sim 5\%$ reflectivity); $E^{(i)}$ - the complex electric field in the i'th fiber near the rear FBG for each fiber laser; $l_i$ - the length of the i'th fiber; $M_{i,j}$ - corresponds to the propagation matrix for a single round trip in the cavity and includes the propagation in each fiber, the output coupler ($\sim 2\%$ reflectivity) and the coupling between the different fibers ($\sim 8\%$ coupling). The light emerging from all the fiber lasers was collected with a lens that was focused onto a detector to obtain the combined output power. Details about the experimental configuration, coupling arrangement and connectivity manipulations were presented in previous work~\cite{Moti25}.}
\end{figure}

Our experiment consisted of an array of 25 coupled fiber lasers schematically presented in Fig.~\ref{system}. Each fiber laser was comprised of a Ytterbium doped double clad fiber with lengths that varied from 1.3m to 1.7m, a high reflecting fiber Bragg grating (FBG) at the rear end of the fiber and a low reflecting FBG at the front of the fiber. Each fiber lasers was end pumped by a stabilized diode laser of $975nm$ wavelength. The length of each fiber laser was about $5m$ and the output wavelength was $1070nm$ with a bandwidth of $10nm$. Accordingly, there were 100,000 available frequencies (longitudinal modes) for each laser. The light emerging from all the fiber lasers was collected with a lens that was focused onto a detector to obtain the combined output power. The fiber lasers were arranged in $5 \times 5$ array, where the coupling between them was achieved by means of four coupling mirrors. By controlling the orientations of the coupling mirrors we could realize a variety of connectivities for the fiber lasers in the array, and in our experiments we concentrated on the one-dimensional and two-dimensional connectivities. Details about the experimental configuration, coupling arrangement and connectivity manipulations were presented in previous work~\cite{Moti25}.

\begin{figure}[h]
\centerline{\includegraphics[width=8cm]{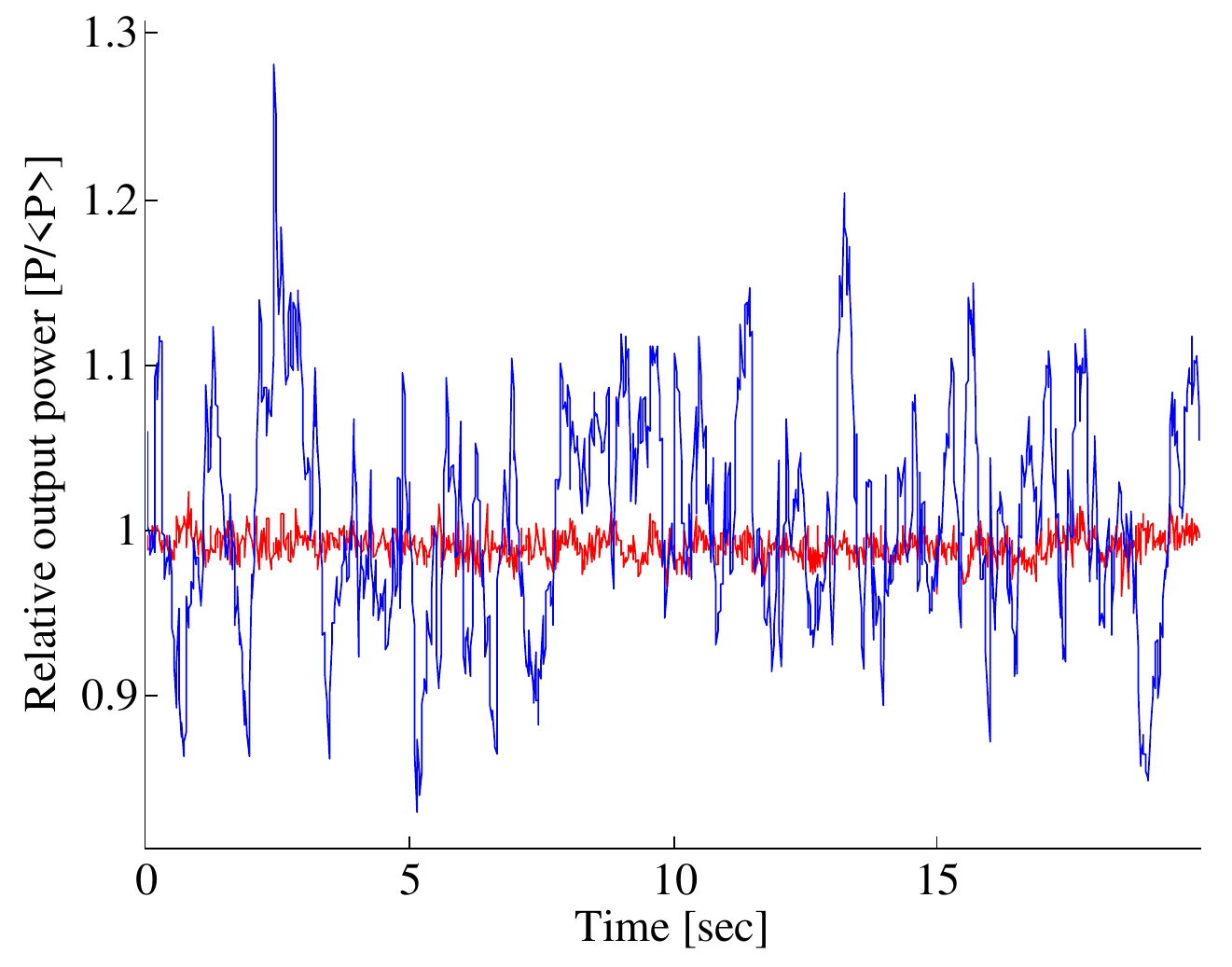}} \caption{\label{bareData} Representative experimental results of the combined output power from the 25 fiber lasers as a function of time. Red curve - without coupling between the lasers; blue curve - with coupling between the lasers. These results indicate that the fluctuations are caused by the coupling between the lasers.}
\end{figure}

The lasers were operated close to threshold to maximize mode competition and ensure that lasing will only occur at the mode where the losses are minimal~\cite{siegman, VarditPRL}. We measured the combined output power from the array over a duration of 60 hours. The correlation time of the output power fluctuations was found to be shorter than $0.5s$, hance we obtained over $500,000$ uncorrelated measurements. Representative results of the combined output power over its mean as a function of time, with and without coupling between the lasers, are presented in Fig.~\ref{bareData}. These are shown over a relatively short time duration, but their behavior was similar throughout the 60 hours measurement. As seen, the power fluctuations with coupling (blue curve) are much larger than those without coupling between the lasers (red curve), indicating that the fluctuations result from the coupling between the lasers~\cite{Galvanauskas16}.

\begin{figure}[h]
\centerline{\includegraphics[width=8cm]{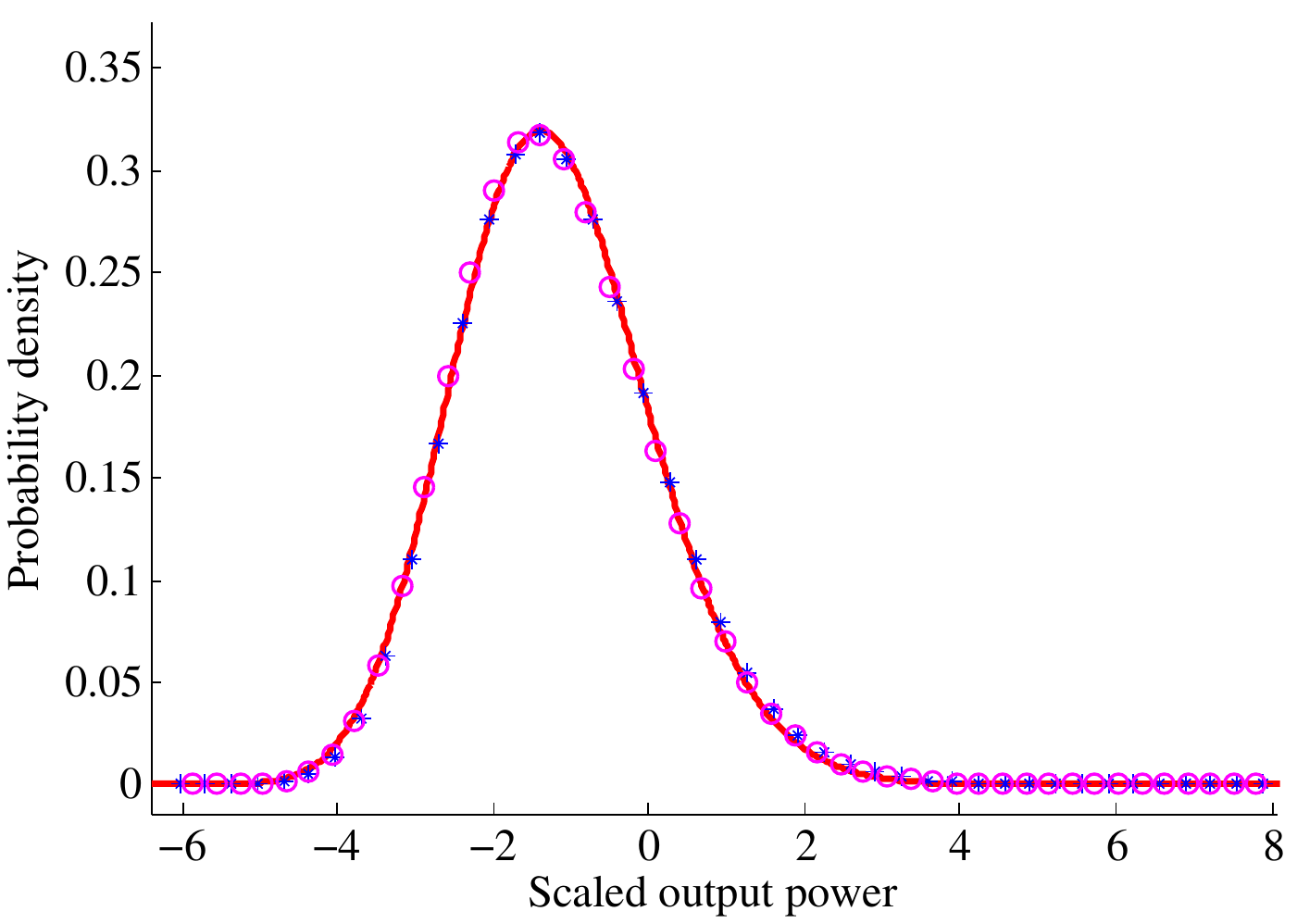}}
\caption{\label{TWlinear} Probability distribution of the scaled combined output power. Circles - experimental results in a one-dimensional connectivity; asterisks - experimental results in a two-dimensional connectivity; solid curve - Tracy-Widom (TW) distribution. As seen, there is a very good agreement between the probability distributions of the experimentally measured results and the TW distribution in linear scale for both connectivities. For closer inspection of the fit to the tails of the distribution see Fig.~\ref{TWM}.}
\end{figure}

Next we compared the measured results to the distribution of the largest eigenvalues of Wishart random matrices. Figure ~\ref{TWlinear} presents the probability distributions of the measured output power in a one-dimensional connectivity (circles) and in a two-dimensional connectivity (asterisks) where the position of the maximum is chosen according to the maximum of the TW distribution (solid curve)~\cite{Edelman}. We present the TW distribution using the scaled units~\cite{Majumdar1}
\begin{equation}
x=\frac{t-4N}{N},
\end{equation}
where $t$ is the largest eigenvalue and $N$ is the matrix size. As evident, there is a very good agreement between the probability distributions of the experimental results and the TW distribution both for the one-dimensional connectivity and for the two-dimensional connectivity.

\begin{figure}[h]
\centerline{\includegraphics[width=8cm]{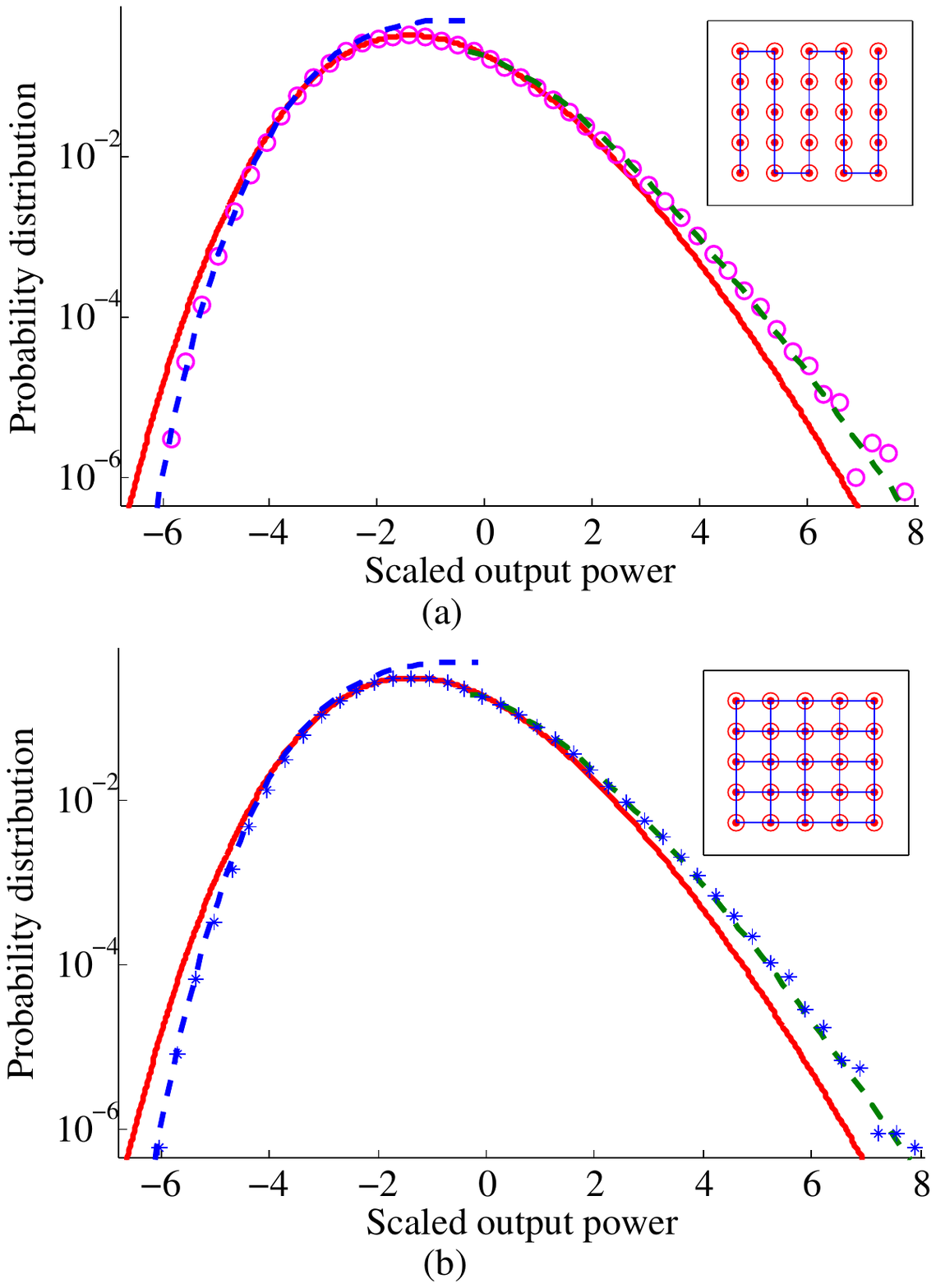}} \caption{\label{TWM}  Probability distribution of the scaled combined output power for (a) One-dimensional connectivity (circles), and (b) two-dimensional connectivity (asterisks) in logarithmic scale. Solid curves - Tracy-Widom (TW) distribution; dashed curves - Majumdar-Vergassola (MV) distribution for eigenvalues much larger (right green) and Vivo, Majumdar and Bohigas distribution for eigenvalues (VMB) much smaller (left blue) than the mean, with no fitting parameters. As seen, at the tails of the measured distribution there are significant systematic deviations from the TW distribution. However, there is a very good agreement between the measured results and the MV distribution both at values which are much larger or much smaller from the mean value, for both connectivities. Insets illustrate the connectivities between the twenty five fiber lasers in each case~\cite{Moti25}.}
\end{figure}

For closer inspection of the tails of the measured distributions, we present in Fig.~\ref{TWM} the probability distributions of the measured combined output power, for one-dimensional connectivity (circles, Fig.~\ref{TWM}(a)) and two-dimensional connectivity (asterisks, Fig.~\ref{TWM}(b)), together with the TW distribution (solid curves) and MV and VMB distributions (dashed curves) on a logarithmic scale. The insets illustrate the connectivities between the 25 fiber lasers~\cite{Moti25}.

The VMB distribution, plotted with no fitting parameters, is~\cite{Majumdar3}
\begin{equation}
P(x)= c_1 \exp \left[ -N^2 \Phi_{-}\left( -x \right)  \right]  \ \textmd{when} \ x < 0 ;
\end{equation}
and the MV is~\cite{Majumdar1}
\begin{equation}
P(x)= c_2 \exp \left[ -N \ \Phi_{+}\left( x \right)  \right]    \ \ \ \textmd{when}  \ x > 0 ;
\end{equation}
where $c_1=0.5$ and $c_2=0.0063$ were found using numerical simulation in ~\cite{Majumdar1}, and the functions $\Phi_{+}(x)$ and $\Phi_{-}(x)$ are
\begin{equation}
\Phi_{+}(x)=\frac{x}{2}+1-\ln(x+4)+\frac{1}{x+4}G\left( \frac{4}{4+x} \right);
\end{equation}
and
\begin{equation}
\Phi_{-}(x)=\ln \left(\frac{2}{\sqrt{4-x}} \right)-\frac{x}{8}-\frac{x^2}{64};
\end{equation}
with $G(z)=\!_3F_2[\{1,1,3/2\},\{2,3\},z]$ a hypergeometric function.

As evident from Fig.~\ref{TWM}, there are significant systematic deviations of the measured distribution from the TW distribution, both at values which are much larger or much smaller than the mean value~\cite{Majumdar1, Majumdar2, Majumdar3}. However, there is a very good agreement between the experimental results and the MV and the VMB distributions for both connectivities, without any fitting parameters. The experimental results of the one-dimensional and the two-dimensional connectivities are essentially identical indicating the universality of the maximal eigenvalue distribution.

In order to illustrate the relation between the distribution of the measured power fluctuations and the maximal eigenvalues of Wishart random matrices, we developed a simple linear model. While an array of coupled fiber lasers is essentially a nonlinear system, many of its properties can be determined by its linearized round trip propagation matrix~\cite{siegman}. For example, the eigenvectors of this matrix correspond to the various global modes of the array while the eigenvalues are $\lambda_n=1-\alpha_n$, where $\alpha_n$ is the loss of mode $n$~\cite{leger}. The tendency of lasers to minimize losses will lead the coupled lasers to operate in the eigenmode corresponding to the largest eigenvalue~\cite{Moti07, VarditPRL}.

We start by letting the electric field $E^{(i)}$ near the rear FBG of each fiber laser be a component of a vector of the total input field $\ket{E_0}$ (see Fig.~\ref{system}). After propagating one round trip, the field $\ket{E_1}$, can be described as $\ket{E_1}=\mathbf{M} \ket{E_0}$, where $\mathbf{M}$ is a $25 \times 25 $ round trip propagation matrix. Details on the derivation of a round trip propagation matrix are presented in  ~\cite{leger}. Specifically, the elements along the diagonal of $\mathbf{M}$ denote the self-feedback light for each laser, as
\begin{equation}
\mathbf{M}_{i,i}=(1-4 \kappa) e^{2 \mathfrak{i}  k_i l_i},
\end{equation}
where $\kappa$ is the coupling strength between two adjacent lasers, $l_i$ the length of the i'th fiber laser and $k_i$ the wave vector of the $i$'th laser out of all the 100,000 available frequencies. The off-diagonal elements of $\mathbf{M}$ denote the coupling between the lasers. For adjacent lasers which are not coupled the corresponding elements are zero. However, when the adjacent lasers are coupled the corresponding elements above the diagonal are
\begin{equation}
\mathbf{M}_{i,j}=\kappa e^{\mathfrak{i}  k_i (l_i+l_j)},
\end{equation}
and below the diagonal the elements are
\begin{equation}
\mathbf{M}_{i,j}=\kappa e^{\mathfrak{i}  k_j (l_i+l_j)}.
\end{equation}

In a resonant cavity at steady state, the vector $\ket{E_1}$ should be one of the eigenvectors of $\mathbf{M}$, so $\ket{E_1}=\lambda_n \ket{E_0}$ where $\lambda_n$ is inversely proportional to the losses in a single round trip. For many round trips, a complex $\lambda_n$ will increase the losses due to interference. These considerations imply that to ensure minimal losses in the combined cavity $\lambda_n$ should be real and maximal~\cite{siegman}. Due to the mode competition between the eigenvectors of $\mathbf{M}$ on the nonlinear gain, the coupled lasers will lase at the mode with the minimal losses~\cite{VarditPRL}, which corresponds to the eigenvector of the round trip propagation matrix with the maximal real eigenvalue. For high gain lasers such as fiber lasers, the output power of a mode is proportional to its eigenvalue~\cite{leger}. Therefore, the combined output power of the array provides a direct measure for the value of the largest eigenvalue of the propagation matrix.

Due to thermal and acoustical fluctuations the length of each fiber laser changes rapidly such that $l_i k_i$ mod $2 \pi$ is effectively a random phase~\cite{Galvanauskas16, FanFluctuations}. These random phases, after each variation in the fiber lengths, result in a different random round trip propagation matrix. The time scale of the length fluctuations in our system is much longer than the relaxation oscillation time of the lasers~\cite{siegman}, justifying the steady sate assumption. Accordingly, the distribution of the combined output power from the array fits the distribution of the largest eigenvalue of random matrices.

The probability for finding a single common wave vector $k$ such that all the lasers in the array will have the same phase and a real valued $\lambda_n$ is negligible --- exponentially proportional to the number of lasers and is $10^{-5}$ for 25 lasers~\cite{no_more_shirakawa, no_more_shakir}. Instead, the lasers group in several clusters where each cluster has its own wave vector~\cite{Strogatz}. Since the light that is coupled from one cluster to the other is essentially lost, the structure of the round trip propagation matrix $M$ is block diagonal, where the elements along the diagonal are $1-4\kappa$ and the off diagonal elements when there is coupling between two specific lasers are $\pm \kappa$. So after each fluctuation we have a different matrix where the blocks sizes and locations and the signs of the off diagonal elements are random. To show that such a round trip propagation matrix $\mathbf{M}$ fall on the Wishart random matrix universality class, we simulated $10^4$ different random realizations of our array with small ($\sim10 \mu m $) fluctuations in the lengths of each fiber. In each realization we found the clusters with common wave vector that yield minimal losses and obtained the corresponding round trip propagation matrix $\mathbf{M}$~\cite{MotiGumbel}. Since $\mathbf{M}$ represents the round trip propagation in the cavity we can define a matrix $\mathbf{X}$ which represents a single pass in the cavity, so, $\mathbf{M}=\mathbf{X}\mathbf{X}^T$. We evaluated the probability distribution of each element in $\mathbf{X}$ and found it to have a Gaussian shape (data not shown), indicating that the round trip propagation matrix $\mathbf{M}$ is indeed a Wishart matrix. Therefore, the distribution of the combined output power from the array should fit to the TW and MV distributions.

To conclude, we presented an experimental configuration of 25 coupled fiber lasers and showed that the probability distribution of their combined output power agrees well with the distribution of the largest eigenvalue of Wishart random matrices, namely the Tracy-Widom, Majumdar-Vergassola and Vivo-Majumdar-Bohigas distributions. We believe that such a configuration can be extended to investigate symplectic and non-Hermitian random matrices with various connectivities by varying the polarizations and the losses in the fiber lasers. Moreover, while in this letter we investigated the combined output power from an array of coupled lasers operating close to threshold, it is possible to operate the lasers far above their threshold and to investigate the phase locking across the array~\cite{Moti25, MotiGumbel}. Such measurement of phase locking gives a direct measure for the number of lasers in each cluster and thereby enables investigation of coupled ensembles of oscillators where a common frequency for all the oscillators can not be found.

\begin{acknowledgments}

This research was supported by the Israeli Ministry of Science and Technology and by the USA-Israel Binational Science Foundation. We are grateful to Eugene Kanzieper for his helpful comments.

\end{acknowledgments}


\end{document}